\documentclass[a4paper,fleqn]{cas-dc}

\usepackage[authoryear]{natbib} % Use natbib with author-year format
\usepackage{xcolor}             % For text coloring



\usepackage{graphicx,times}
\usepackage{subcaption}
\usepackage{amssymb,amsmath}
\usepackage{aas_macros}

\usepackage{soul}
\let\oldAA\AA
\renewcommand{\AA}{\text{\normalfont\oldAA}}
\usepackage{breakcites}

\def\tsc#1{\csdef{#1}{\textsc{\lowercase{#1}}\xspace}}
\tsc{WGM}
\tsc{QE}
\begin{document}

%%%%%%%%%%%%%%%%%%%%%%%%%%%%%%%%%%%%%%%%%%%%%%%
\let\WriteBookmarks\relax
\def\floatpagepagefraction{1}
\def\textpagefraction{.001}

\shorttitle{HEP-tail generation mechanism for XRBs}    

\shortauthors{Kumar N}

%\title [mode = title]{Multi-wavelength temporal and spectral study of the FSRQ B2\,1348+30B}  
\title [mode = title]{High energy power-law tail in X-ray binaries spectrum and bulk Comptonization due to a conical outflow from a disk } %% Article title

\tnotemark[<tnote number>] 

%\tnotetext[<tnote number>]{<tnote text>} 


\author[1,2]{Nagendra Kumar}

\cormark[1]

%\fnmark[<footnote mark no>]

\ead{ngendra.bhu@gmail.com}

%\ead[url]{<URL>}

%\credit{<Credit authorship details>}

%% \begin{frontmatter}


%% \author[1,2]{Nagendra Kumar} %% Author name

\affiliation[1]{organization={Department of Physics, Indian Institute of Science},%Department and Organization
         %  addressline={}, 
            city={Bangalore},
            postcode={560012}, 
         %  state={},
            country={India}}
 \affiliation[2]{organization={Aazad Path, New Bengali Tola, Mithapur},
         %   addressline={},
             city={Patna},
             postcode={800001},
          %   state={},
             country={India}}

 %% Abstract
\begin{abstract}
X-ray binaries (XRBs) often exhibit a high energy power-law tail (HEP-tail) and
these tails can be generated by the bulk
Comptonization (BMC) process with a free-fall bulk region onto the compact object. The radio emission (which is generated by a synchrotron-emitting outflowing electrons) is observed in all spectral state of XRBs. Interestingly, the variations of HEP-tail flux among different spectral states is similar to the variation of radio flux.  We motivate to study the HEP-tail in BMC process with an outflowing medium.
For this we consider a collimated and conical (of opening angle $\theta_b$ with axis
perpendicular to the accretion disk) outflow geometry. We simulate the BMC spectrum by using a Monte Carlo scheme.   
We find that the emergent spectrum has power-law tail (of
photon index $\Gamma$ $>$ 2 and with high energy cut-off $E_{c}$ $>$ 200 keV)
only for $\theta_b$ greater than $\sim$30 degrees in conical outflow, 
while for a collimated or a conical outflow ($\theta_b$ $<$ 30 degrees)
these HEP-tail can be only generated when it is also found in thermal
Comptonized spectra (i.e., at sufficiently high Comptonizing medium temperature). These results are approximately consistent with analytically derived expressions.
We describe the observed GRS 1915+105 spectrum for two classes $\chi$ and $\gamma$ in conical outflow, for this the outflow speed is highly relativistic and the kinetic power of wind suggest that the HEP-tail can be generated at inner region of the accretion disk, like inner disk radio emission.
  
\end{abstract}

%%%%\begin{graphicalabstract}
%%%%\end{graphicalabstract}

%%%%%%\begin{highlights}
%%%%%%\item Research highlight 1
%%%%%%\item Research highlight 2
%%%%%%\end{highlights}

%% Keywords

%% \end{frontmatter}

\maketitle

%%%%%%%%%%%%%%%%% BODY OF PAPER %%%%%%%%%%%%%%%%%%
\section{Introduction}
 
Black hole (BH) low mass X-ray binaries (LMXBs) frequently change their X-ray spectral
states 
particularly power-law dominated low intense hard (LH) state to black body
dominated high intense soft (HS) state via outburst, which again settle
down to the LH state.
During the spectral state transition, sometimes a very high intense power-law dominated (VHS) state with
photon index $\Gamma$ greater than 2.4 is observed without an exponential high energy cut-off ($E_c$)  and generally referred 
as a steep power law (SPL) state
(\citealp[for a review see,][]{McClintock-Remillard2006, Done-etal2007}). 
A high energy power-law tail (hereafter HEP-tail) is observed in both states LH and HS with $\Gamma$ $>$ 2.0, which usually extends up to $\geq$ 200 keV (or $E_c >$ 200 keV) 
\cite[e.g.,][]{McConnell-etal2002, Motta-etal2009, Titarchuk-Shaposhnikov2010},
and it is also detected in the neutron star (NS)
LMXBs \cite[e.g.,][]{Revnivtsev-etal2014}.

The origin of the low energy (2-100 keV) power-law component has been thought as a thermal Comptonization (TC) of the disk photons, but 
for the HEP-tail  a modified version of TC has been invoked. For example, \citet{Done-Kubota2006} assumed
that the disk 
and Comptonizing medium are energetically coupled to each other
\cite[see also,][]{Kubota-Done2016}; in the so-called  hybrid model an hybrid distribution of electron velocities with a  thermal plus
a non-thermal origin [or power law], has been proposed
\cite[][]{Coppi-1999, Gierlinski-etal1999} and in  the
bulk Comptonization (BMC) model the electrons are in a free-fall converging flow of
spherically accreted plasma into the BH (\citealp{Titarchuk-etal1997}, \citealp[and for NS e.g.,][]{Farinelli-etal2009}). 

In the BMC framework, for a spherically diverging outflow, \citet{Laurent-Titarchuk2007} had 
noticed only downscattering of the soft spectrum, i.e., the HEP-tail has not been generated in this model
\cite[see also,][]{Psaltis-2001, Ghosh-etal2010}.
However, the outflow is observed in all spectral states of LMXBs, e.g., the radio jet outflow is in LH $\&$ SPL states, and wind outflow is in HS state
(\citealp[e.g.,][]{Fender-Belloni2012, Ponti-etal2012, Miller-etal2016a, Trigo-Boirin2016, Degenaar-etal2016}).
Wind and jet can  exist simultaneously e.g., in LMXBs \cite[][]{Homan-etal2016, Drappeau-etal2017, Tetarenko-etal2018}, in active galactic nuclei \cite[e.g.,][]{Tombesi-etal2014}. Morphologically, in some systems the inferred wind outflow geometry has a conical shape \cite[e.g.,][]{Knigge-etal1995, Tombesi-etal2015, Degenaar-etal2016}.
Although the radio emission is mainly associated to the LH and SPL states,  a fainter radio emission has been observed also in the HS state. 
In general the radio emission is generated at the inner accretion disk by synchrotron emitting  outflowing relativistic electrons \cite[e.g.,][]{Fender-etal1999, Fender-etal2004, Fender-etal2009,  Fender-Hendry2000, Muno-etal2001, Fender-Kuulkers2001,  Rushton-etal2010}.
%%%
Interestingly, the HEP-tail
%high energy power-law tail
flux in the HS state is also comparatively smaller than in the VHS state.
Also, the jet power of the VHS is positively correlated with the peak X-ray luminosity of the HS \cite[e.g.][]{Zhang-Yu2015}.
It might be possible that the emission mechanism of the HEP-tail is related, in some sense, to the emission mechanism for radio emission.
In this work, we attempt to describe the HEP-tail emission mechanism in outflowing medium in the BMC framework.
%In this paper, we motivate to study the high energy power-law tail by BMC process in outflow with a conical and collimated outflow.
For outflowing medium we consider two different geometries,  conical and collimated outflow.
We compute the spectrum by using a Monte Carlo method for BMC.
We find that when the outflow is a conical type  (i.e., at any scattering point P the electron's outflow/ bulk direction is any one of the directions inside the cone of opening angle $\theta_b$ at P) then the soft
photons can get upscattered and the emergent spectrum has a power-law
component.

%
%
%
% 
%\section{Monte Carlo Method}
\section{Calculations and Results}\label{sec:cal-res}
{\tt \textbf{Method :}} In bulk Comptonization, the photons are upscattered due to
both thermal and bulk motions of the electrons and it was %initially
formulated by \citet{Blandford-Payne1981a, Blandford-Payne1981b}.
The average energy exchange per scattering $\Delta E$ for a monochromatic photon
of energy $E$ in the Comptonizing medium (or corona) of temperature $T_{e}$ and   constant bulk speed $u_b$ (particularly, it is derived for the converging flow of thermal plasma, in which the outward diffusing photon mostly experienced a head-on collision with the inflowing electron, \citealp[e.g.,][and references therin]{Titarchuk-etal1997}) 
is
\begin{equation}\label{delta_E_bmc}
  \Delta E = \Delta E_{TH}+ \left( \frac{4u_b}{c\tau} + \frac{4(u_b/c)^2}{3} \right) E \end{equation}
Here, $\tau$ is the optical depth of the scattering medium, $m_e$ is the rest mass of the electron, $k$ is the Boltzmann constant.
$\Delta E_{TH} = (4kT_e - E)\frac{E}{m_e c^2}$ is for thermal Comptonization (i.e., $u_b$ = 0).  The second and third right-hand side terms of equation (\ref{delta_E_bmc}) are the Comptonization due to the first and second order in bulk velocity of electron, respectively. % and the second term is for the random scattering. And}
  %For a considered Comptonizing geometry in a converging flow of thermal plasma by \citet[][references therein]{Titarchuk-etal1997} the first order term in $u_b$ of $\Delta E$ is attributed dominantly by the head-on Compton scattering (see also, appendix \ref{head-on}).
The first order term in $u_b$ of $\Delta E$ is dominated %associated dominantly
by the head-on Compton scattering (see also, appendix \ref{head-on}), so this term, in general, is applicable for any medium where the electrons bulk velocity is opposite to the photons direction.
  The second order in $u_b$ of $\Delta E$ is analogous to the increase in the scattered photon energy by the electron thermal motion \cite[e.g.][]{Psaltis-lamb1997},  i.e., it is attributed by the isotropic distribution of Compton scattering or random Compton scattering.

 Since, the time averaged Comptonized spectrum depends on $\Delta E$ and the average number of scatterings $\langle N_{sc}\rangle$ \cite[e.g.][]{Kumar-Misra2016}.
 Hence from equation (\ref{delta_E_bmc}) with neglecting the first order term in $u_b$, in general for a given $\langle N_{sc}\rangle$ (or $\tau$), the BMC spectrum would  degenerate over $kT_e$ and $u_b$. That is for a given spectrum  at fixed $\langle N_{sc}\rangle$, %the many sets of $(kT_e, u_b)$ are possible that give same $\Delta E$.
there are many possible sets  of BMC parameters  $(kT_e, u_b)$ that give same $\Delta E$ including the TC. %parameters (i.e., $(kT_e, u_b = 0)$)}.
As for a special case the analytic solution for TC is known [e.g., \citealt[][and references therein]{Kumar-Misra2016}]), we will verify the calculations by comparing the BMC results with TC which has the same $\Delta E$ as BMC (see Figure \ref{chk-delE}). % using it with considering one of set $(kT_e, u_b=0)$ for TC (see Figure \ref{chk-delE}.
We calculate the spectrum using a Monte Carlo (MC) method for BMC. We adopt a more general approach to develop the MC algorithm, where we assume that the electron has two parts of velocity, one is the bulk velocity and another one is the random oriented thermal velocity. The MC method has been described in appendix \ref{MC-method}.

\begin{figure}\hspace{-0.5cm}
\centering
\includegraphics[width=0.45\textwidth]{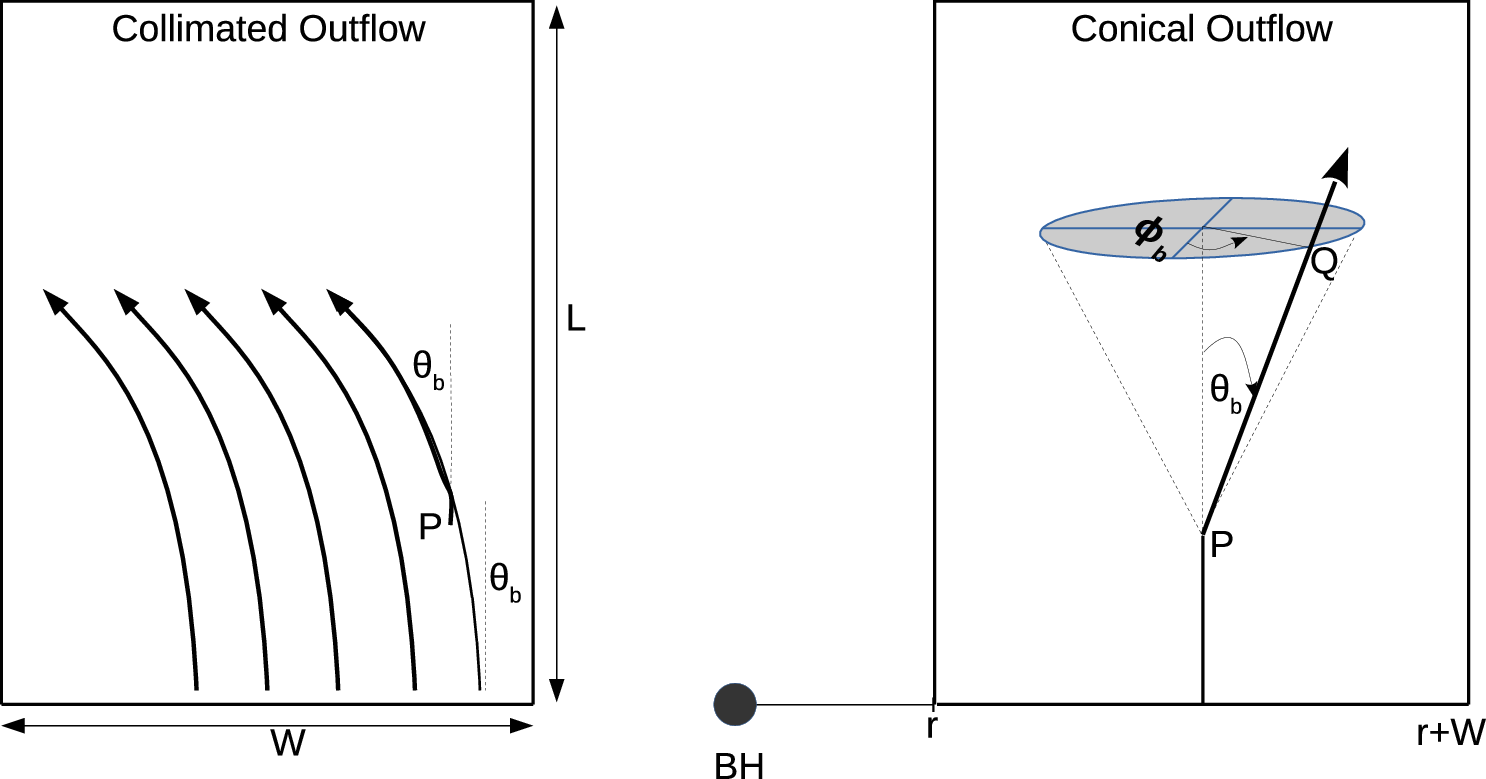} \vspace{-0.20cm}
\caption{Meridional cut of a rectangular torus  shaped Comptonizing medium (or corona) of width W and height L  surrounding the BH to study the outflow motion. In the left side, a schematic diagram for the collimated flow of angle $\theta_b$  is presented. In the right side we show a schematic diagram for the conical flow of opening angle $\theta_b$. Here,
the conic shaded region represents a possible conical
outflow direction at the scattering point P, and the PQ arrow indicates an arbitrary outflow direction ($\theta_b$, $\phi_b$) in  local coordinates.
}
\label{geo-outflow}
\end{figure}

\begin{figure*}
\centering$
\begin{tabular}{lcr}\hspace{-0.9cm} 
  \includegraphics[width=0.36\textwidth]{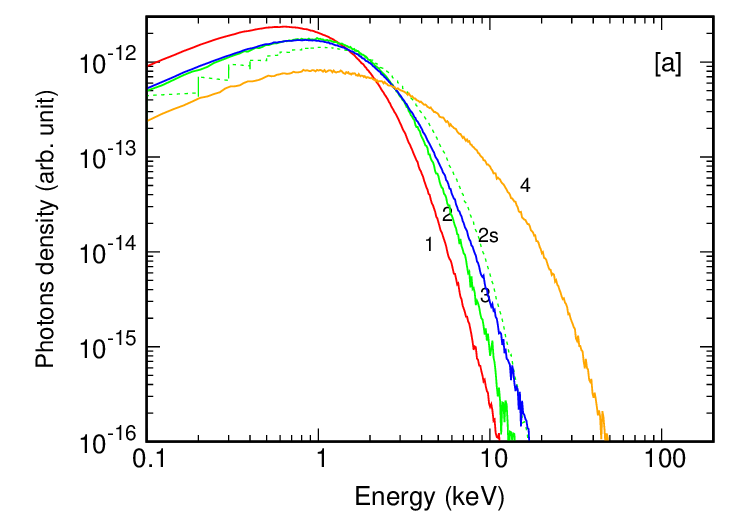} &\hspace{-0.8cm}
  \includegraphics[width=0.36\textwidth]{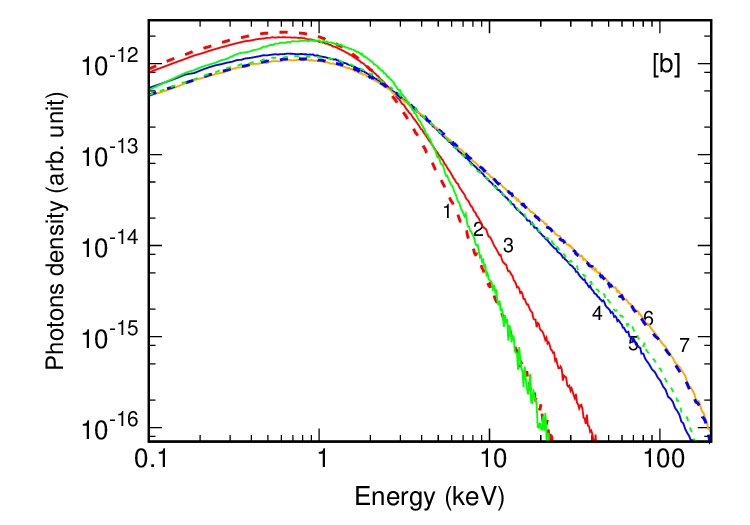} &\hspace{-0.8cm}
  \includegraphics[width=0.36\textwidth]{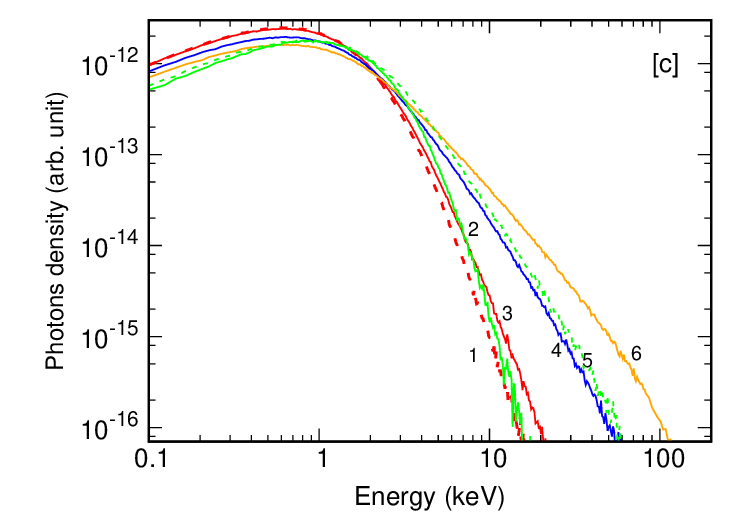} \\
\end{tabular}$
\caption{The simulated emerged spectra for different outflows geometry for a low medium temperature.
The left is for collimated outflow, while the middle and right panels are for the
case I and case II of conical outflow, respectively (see text).
In the left panel the curves 1 and 2 are for $\theta_b$ = 0 (outflow) and 180
(inflow) degrees, respectively, and for the curves 3 and 4, the $\theta_b$ is 90 degrees. The curve 2s is for the single scattering at $\theta_b$ = 180 degrees.
In the middle and right panels the curves 1,2,3,4,5
and 6 are for $\theta_b$ = 20, 160, 30, 60, 110 and 90 degrees, respectively.
In the middle panel the curve 7 (dashed) is for 95 degrees. The spectral
parameters for all
curves are $kT_e$=3.0 keV, $kT_b$=0.5 keV, $\tau$ = 3 and
$u_b$ = 0.45$c$ except for the curve 4 of left panel where $\tau$ = 15
and $u_b$ = 0.65$c$.}
\label{spec-outflow-ang}
\end{figure*}

\begin{figure}
\centering$
\begin{tabular}{lr}\hspace{-0.90cm} 
  \includegraphics[width=0.28\textwidth]{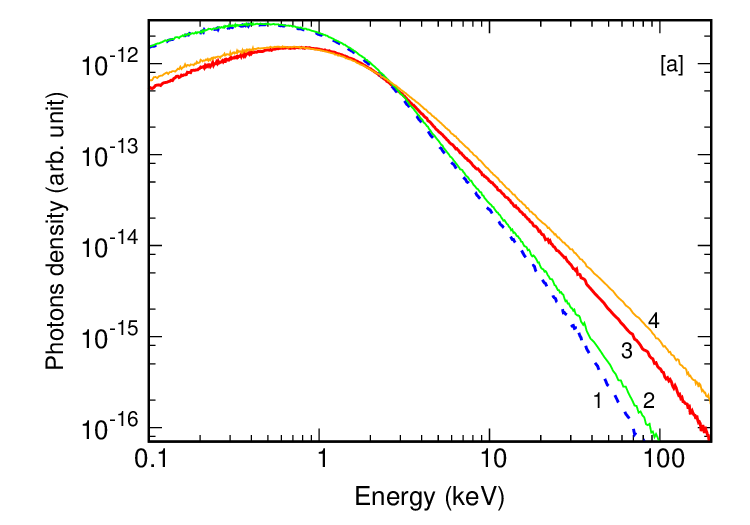} &\hspace{-1.cm}
  \includegraphics[width=0.28\textwidth]{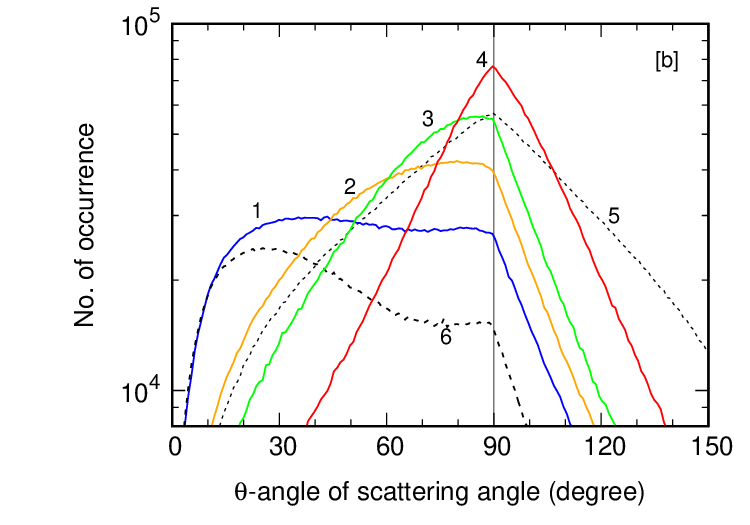} \\ 
\end{tabular}$
\caption{The emergent BMC spectra (left panel)  and the distribution of scattering angle (right panel).  
    The left panel is for an emergent spectra for a high medium temperature
($kT_e$ = 30 keV).
The curve 3 is for a spectrum of TC
dominated case ($(u_b/c)^2$ $\ll$ $3kT_e/(m_ec^2)$), while others are for
bulk dominated case, $u_b$ = 0.75c.
The curve 1 is for collimated outflow of $\theta_b$ = 15 degrees, and the
curves  2 and 4 are for conical outflow of $\theta_b$ = 15 and 45 degrees,
respectively. 
The right panel is for scattering angle distribution. 
Curves 1, 2, 3 and 4 are for conical outflow of $\theta_b$ =
15, 45, 60 and 90 degrees, respectively and $u_b$ = 0.75c, and the curve 6 is for
$\theta_b$ = 15 degrees and $u_b$ = 0.85c. The curve
5 is for TC dominated case. The rest parameters are $kT_b$ = 0.5 keV and
$\tau$ =  1.}
\label{spec-td-ud-ang}
\end{figure}

{\tt \textbf{Calculation geometry :}} We assume that the outflow occurs on the accretion disk from the radius $r$ to $r+W$, %for
so the outflow region will be a torus shaped with a  rectangular cross-section of width $W$ and height $L$.
%surrounding the compact object, which has %a \cite[e.g.,][]{Kumar-Misra2016}.
The meridional cut of the torus is shown in Figure \ref{geo-outflow}. 
We fix a global spherical polar coordinate (r, $\theta$, $\phi$) at the
center of the torus (or at the compact object). Without loss of generality we assume that  the torus exists only above the equatorial plane. %
The Compton scattering takes place inside the torus, which we also refer as a  Comptonizing medium or a corona.
We assume that the optical depth $\tau$ is along the vertical direction, and when the scattered photons cross the equatorial plane, these photons will get absorbed.
%.
The seed photon source is a black body at temperature $T_b$. It emits vertically from the equatorial plane of the accretion disk, and situates inside the torus, which also asserts that the BMC calculation does not depend on whether the torus  lies above the disk at some height or not. 
Depending on the $\frac{L}{W}$ ratio and the mean free path of the photons inside the torus ($\lambda$, and here $\lambda \ll W$), the scattered photons can escape either from the top of the torus (1st case) or from the outer boundary of the torus (2nd case) or both. The 1st case dominates for $\frac{L}{W} \ll 1$, which is more relevant for XRBs.
%The 1st case is more relevant, in general, for XRBs,
Hence, we consider only the 1st case for the calculation with $\frac{L}{W}$ = 0.1.

For the outflow geometry we consider two physical situations (a) collimated outflow, and (b) conical outflow.
In collimated outflow, the electron's medium flows in $\theta_b$ direction (i.e., away from the compact object) with bulk speed $u_b$ at any scattering point P on fixed $\phi$-plane in the global coordinate as shown in the left side of Figure \ref{geo-outflow}.
In conical outflow, to assign the outflow direction we consider a cone of opening angle $\theta_b$ with axis perpendicular to the equatorial
plane at scattering point P.
The outflow direction will be  %any one of possible direction
any of the possible straight lines
inside the cone from the P as shown in the right side of Figure \ref{geo-outflow}.
The torus temperature is $T_e$. Inside the torus the electrons execute simultaneously thermal motion and bulk motion. The bulk motion is generally referred a movement in uni-direction, e.g., here collimated flow. However in the conical outflow case, the medium as a whole moves in upward direction but the electron's bulk motion direction is not unique as shown in the right side of Figure \ref{geo-outflow}. For the first scattering the $\Delta E$ for both outflow geometries is same as the angle between the incident photon and the bulk velocity of electron is fixed to $\theta_b$, %incidents in vertically upward direction,
  and its averaged  magnitude is determined as (see the appendix \ref{head-on})
\[\frac{\langle \Delta E \rangle}{E} 
  = \frac{\frac{u_b}{c} (-\cos\theta_b +\left(\frac{16}{9\pi}\right)^2\frac{u_b}{c} \sin^2\theta_b)}{1-\left(\frac{u_b}{c}\frac{16}{9\pi}\sin\theta_b\right)^2}   \]
      
%
%
%%%%%
\subsection{Results}
To explore the general characteristics of the BMC spectrum we study a wide range of $\theta_b$ for both collimated and conical outflows. As our primary aim is to explore the HEP-tail of XRBs observed in all spectral states, we consider a
typical mid value of the seed photon source temperature $kT_b = 0.5$ keV.
%corresponding to all states.
We take two different corona temperatures $kT_e$ = 3 and 30 keV, in which the low value may characterize the HS state and high value is for the LH state \cite[e.g.][and references therein]{Kumar-Misra2016a}. As discussed, the TC fails to generate the observed HEP-tail, %also our interest is in BMC spectrum,
we consider only the bulk motion dominated case or $(u_b/c)^2$ $\gg$ $3kT_e/(m_ec^2)$, and take $u_b$ = 0.45 and 0.75$c$ for  low and high $kT_e$, respectively.
As $\Delta E$ increases with increasing $T_e$, we consider $\tau$ = 3 and 1 for  low and high $kT_e$, respectively.
Particularly, to determine the outflow direction for the conical outflow geometry we consider two different plausible cases.
In case I we simply take, the outflow direction at scattering point P is in
any one of directions on the surface of the cone from its vertex P,
that is, $\theta_b$ = constant
and $\phi_b$ varies from 0 to 2$\pi$ (see right side of Figure \ref{geo-outflow}).
In case II, the outflow direction can be any one of directions 
within the conical region 
so here $\theta_b$ will vary from 0 to $\theta_b$ = constant,
and $\phi$ will vary again from 0 to 2$\pi$.

{\tt \textbf{Emergent spectrum from Outflow geometry:}}
The emergent spectra at low $kT_e$ for the collimated outflow are shown in Figure \ref{spec-outflow-ang}a for three different $\theta_b$ = 0 (curve 1), 180 (curves 2 and 2s) and 90 (curves 3 and 4) degrees. In which the curve 2s is for the single scattering and curve 4 is for the $\tau$ = 15 and $u_b$ = 0.65c.  
We do not find a high energy power-law tail in the collimated outflow
even in the extreme condition (higher $\tau$ and larger $u_b$) like curve 4 of Figure \ref{spec-outflow-ang}a. This result is consistent with the analytic one, where the magnitude of $\Delta E$ ($ > $0) is almost zero (see the appendix  \ref{head-on}).  % at low medium temperature ($kT_e$= 3 keV).
Since, the seed photons direction is vertically upward, so in the first scattering the photon will collide head-on with the electron  for $\theta_b$ =180 degrees, and for $\theta_b$ = 0  it will hit the electron from behind. As a result the single scattered spectrum for $\theta_b$ = 180 degrees is harder than the spectrum for $\theta_b$ = 0.
The Klein-Nishina cross section predicts that for relativistic electrons the scattered photon will align in the incident electron direction (e.g., see Figure \ref{spec-td-ud-ang}b). That is, for the next scattering the photon will hit the electron from behind, therefore for multiple scattered BMC spectrum for $\theta_b=180$ degrees would be red-shifted from the single scattered spectrum. We find the same result, which is shown by the curves 2 and 2s in Figure \ref{spec-outflow-ang}a \cite[see also][]{Janiuk-etal2000}.

In Figure \ref{spec-td-ud-ang}a the curve 1 is for the emergent spectrum at large $kT_e$ for collimated outflow with $\theta_b$ = 15 degrees and $u_b$ = 0.75$c$. We notice a power-law tail in the spectrum. To confirm this 
we also compute the thermal Comptonized spectrum (or TC dominated
spectrum i.e., $(u_b/c)^2$ $\ll$ $3kT_e/(m_ec^2)$)  at same $kT_e$= 30 keV, which is shown by the curve 3.
We find that in this case the collimated BMC spectrum for any $\theta_b$ is always softer (or downscattered) in comparison to TC dominated one.  
Since for a given $\phi$-plane the spherical diverging outflow will appear as a collimated outflow with varying $\theta_b$ from 0 to 90 degrees above the disk, so the results are consistent with those of \cite{Laurent-Titarchuk2007}. 
Conclusively, in collimated
outflow the power-law tail can be found for sufficiently large $kT_e$ at where TC dominated spectrum has also power-law tail.
The above results are consistent with \citet{Kylafis-etal2014},
where they described the power-law component in soft gamma-ray
repeaters by BMC process with having a vertically downward bulk region onto the neutron star pole for a large kT$_e$. %
\cite[see also ][for a power-law component of gamma-ray burst spectrum]{Titarchuk-etal2012}.

Next, we  compute the emergent spectrum of conical outflow at low $kT_e$, the results are shown  in Figure \ref{spec-outflow-ang}b and  Figure \ref{spec-outflow-ang}c for case I and case II, respectively. 
In general, we find that the photon index decreases with $\theta_b$ for both cases, and the spectrum of case I is harder than the case II.
Interestingly we find that to generate the power-law tails with $E_c$ $>$ 200 keV
and observed range of
$\Gamma$ ($>$ 2), the $\theta_b$ should be greater than 30 degrees and $u_b$ $>$
0.4c (for low $kT_e$).
Evidently, the case II is a more plausible case for conical outflow in the system, %, and the BMC spectra are similar (slightly different) in both cases.
however, for simplicity to extract the general picture, we consider the case I only.  %however to extract the general picture for BMC in conical outflow we consider the case I as the spectra is slightly different for both cases but trend is same.
For large $kT_e$ (= 30 keV), the emergent spectrum for case I with $u_b$ = 0.75$c$ is shown in the Figure 
\ref{spec-td-ud-ang}a. Here curves 2 and 4 are for $\theta_b$ = 30 and 45 degrees, respectively.
We find that the general trend of results is similar to the collimated case, i.e., when there is no HEP-tail for a given $\theta_b$ at low $kT_e$ then at large $kT_e$ their spectrum is softer than the TC dominated spectrum. 
As we note here that  
for $\theta_b$ $<$ 30 degrees, the $\Gamma$ is higher from TC dominated one, 
while the opposite is true for the case of $\theta_b$ $>$ 30 degrees.
%.

Next to understand this distinction in details which occurs around $\theta_b =$ 30 degrees, we compute the distribution 
of scattering angle (the angle between scattered and incident photon)  for different values of $\theta_b$ in the conical outflow (case I). The results are shown  in the Figure \ref{spec-td-ud-ang}b.
%in Figure \ref{spec-td-ud-ang}b, the distribution 
%of scattering angle (the angle between scattered and incident photon)
%has been shown for a conical outflow (case I).
Here,  the curves 1, 2, 3 and 4 are  for $\theta_b$ = 15, 45, 60 and 90 degrees, respectively with $u_b = 0.75c$ and curve 6 is for $\theta_b$ = 15 degrees and $u_b = 0.85c$. The curve 5 is for TC dominated case. We find that the
scattered photons are in the outflow direction of the electrons and it tends to be more 
along the outflow direction for a large value of $u_b$ (see curve 6).
Expectedly, in terms of the randomness
the outflow direction becomes more random with increasing $\theta_b$ and it becomes
completely random (like TC ones) when $\theta_b$ tends to 90 degrees. We find a similar correspondence in terms of scattering angle, see the curves 4 and 5. Thus, it is expected here that after some value of $\theta_b$ the BMC spectrum becomes harder in comparison to the TC dominated one. Analytically, we show that the approximate $\Delta E$ for conical flow is positive and its magnitude increases with increasing $\theta_b$ (see appendix \ref{head-on}), which confirms the reliability of MC calculations.
We find here that for $\theta_b$ $>$ 30 degrees the $\Gamma$ is smaller in comparison to the TC one, and this holds almost upto $u_b = 0.99c$. 
%and for $\theta_b$ $<$ 30 degree there is no power-law tail in spectra due to mainly downscattering.
Hence, in general, as the collimated case in a conical outflow for $\theta_b$ $<$ 30 degrees, the
power-law tails can be produced when the power-law tails are also produced in
TC dominated case.
%Since, \re{the} BMC spectrum is degenerate over $kT_e$, $u_b$ and $\tau$, \re{so} to illustrate this, in Table \ref{obs-comparison} we list the few ranges of these parameters

As discussed earlier, the BMC spectrum degenerates over $kT_e$, $u_b$ and $\tau$. To demonstrate this degeneracy, we list a few sets of these parameters in Table  \ref{obs-comparison}
for a fixed typical 
photon index $\Gamma \sim$ 2.5 for case I of conical outflow along with $E_c$.
We find, in general, that the $E_c$ increases with $u_b$ for a given $kT_e$.

\begin{table}
\caption{Sets of (outflow speed $u_b$, optical depth $\tau$) and 
high energy cut-off $E_{c}$ of bulk dominated BMC spectra of photon index $\Gamma$
$\sim$2.5 in conical outflow.}
%\begin{ruledtabular}
\center
\label{obs-comparison}
\begin{small}
\begin{tabular}{p{1.cm} p{2.cm}  p{2.cm} p{2.cm} }
\hline 
 & \multicolumn{3}{p{7.2cm}}{\hspace{1.6cm} ($u_b$, $\tau$), $E_{c}$ $^a$ (MeV)
when } \\ \hline%{2-4} 
kT$_e$= &$\theta_b$=30$^o$ & $\theta_b$=45$^o$ & $\theta_b$=60$^o$ \\ \hline
3keV & (0.8c, 5.2), 0.7 & (0.72c, 2.0), 0.6 & (0.6c, 1.9), 0.4 \\
 & (0.85c, 4.2), 0.8 &   (0.85c, 1.2), 0.8 &   (0.7c, 1.2), 0.6 \\ \hline
30keV & (0.8c, 2.8), 1.1 & (0.85c, 0.8), 1.2 & (0.7c, 0.6), 0.8 \\
& (0.7c, 2.6), 0.8 &   (0.7c, 0.9), 0.8 &   (0.6c, 0.7), 0.6 \\ \hline

%\end{ruledtabular}

\multicolumn{4}{p{8.4cm}}{{\footnotesize $^a$ here, $E_{c}$ increases with $u_b$ (or $kT_e$)  for given $kT_e$ (or $u_b$); and for fixed value of $E_{c}$ and $\theta_b$, $u_b$ and $\tau$ decrease by increasing $kT_e$.}}
\end{tabular}\\[0.2cm]\end{small}
%\vspace{-0.25cm}
\end{table}

\begin{figure}\vspace{-2.5cm}
%\hspace{2.0cm}
  \includegraphics[width=0.55\textwidth]{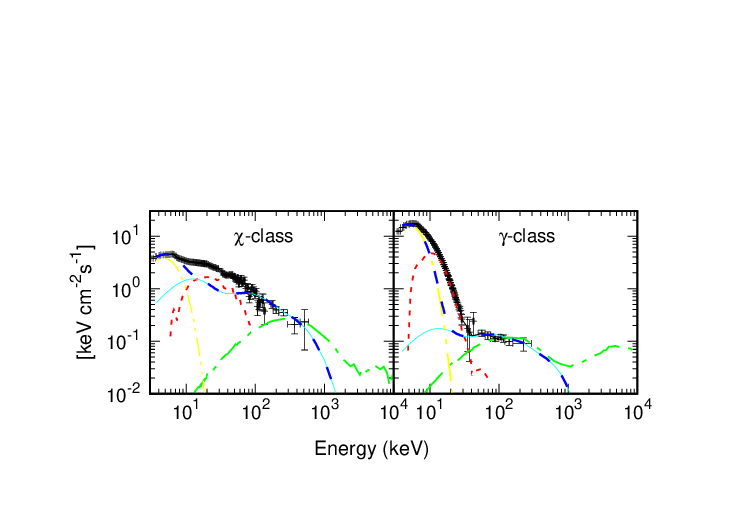}
\vspace{-1.20cm}
\caption{Comparison of the $\chi$ (left panel) and $\gamma$ (right panel) classes
of GRS 1915+105 with the bulk Comptonization model, here the data points are
taken from \citet{Zdziarski-etal2001}. The two-dotted-dashed, solid, dashed, and
dotted curves are for 
black body (BB), bulk Comptonization (BMC), BB + BMC, and residual (data-(BB+
BMC) components, respectively. Here, $\theta_b$ = 30 degrees, $u_b$ = 0.95c,
$kT_e$= 3 keV, $kT_b$ = 1.2 keV and $\tau$ = 2.0 ($\chi$-class) = 2.8 ($\gamma$-class). Dotted-dashed curve is for BMC component when $u_b$ = 0.998c and $kT_e$ = 100keV, here the other components are not shown for clarity.}
\label{result-compare1}
\end{figure}

{\tt \textbf{Comparison with observations:}} In Figure \ref{result-compare1} we compare qualitatively the observed HEP-tail of $\chi$ (LH) and $\gamma$
(HS) classes 
of GRS 1915+105 with bulk Comptonized spectra  at two values of $kT_e$ = 3 and 100 keV for a conical outflow geometry of case I.
The data points are taken from \citet[][see theirs Fig. 2]{Zdziarski-etal2001}, the observation ID for $\chi$ and $\gamma$ is VP 619 and VP 813, respectively. %, see also theirs model parameters value.
The computed BMC spectrum for both classes are shown by cyan solid and green dotted-dashed curves for low and high $kT_e$, respectively.
We find that the BMC spectrum well describe the HEP-tail for both classes for both cases of $kT_e$.
For completeness of spectral modeling, for low $kT_e$ we also show the black body (BB, shown by yellow two-dotted-dashed curve), BB+BMC (blue big-dashed curve) and [data points - (BB + BMC)] (residual, shown by red short-dashed curve). In general the residual can be described either  due to a relativistic reflection 
spectra \cite[e.g.,][]{Ross-Fabian2005} or due to TC component or both \cite[e.g.,][]{McConnell-etal2002, Farinelli-etal2009, Revnivtsev-etal2014, Kubota-Done2016}.
The spectral parameters set for low $kT_e$ are $\theta_b$ = 30 degrees, $u_b$ = 0.95$c$, $kT_b$ = 1.2 keV and $\tau$ = 2.0 ($\chi$-class),  2.8 ($\gamma$-class). For high $kT_e$, the parameters are  $\theta_b$ = 30 degrees, $u_b$ = 0.998$c$, and ($kT_b, \tau$) = (0.85 keV, 0.8) (for $\chi$-class), (0.35 keV, 3.0) (for $\gamma$-class).

As the HEP-tail is generated by a BMC process with conical outflow geometry, like leptonic case of blazar radio jet, we assume that the hadrons do not contribute in the radiative process, the wind mechanical power is significantly due to the electrons, and the $P_{wind}$ is \cite[e.g.,][]{Zdziarski2014} $P_{wind}$ = $\dot{M}_{wind} c^2 \big(\frac{1}{\sqrt{1-(u_b/c)^2}}\big)$, here $\dot{M}_{wind}$ is the mass outflow rate $\dot{M}_{wind}$ = $\Omega\mu m_e n_e R^2 u_b$,  $\Omega$ 
%%$(= 2\frac{2\theta_b}{4\pi}\big)$
is the covering factor ($0 < \Omega < 4\pi $, here $\Omega = (2\theta_b4\pi)/360 $, and $\theta_b$ is in degrees), $\mu$ is the mean atomic weight,
$m_e$ is the mass of the electron, $R$ is the launching radius and $n_e$ = $\frac{\tau}{L\sigma_T}$ is electron density.
For low $kT_e$, $\dot{M}_{wind}$ =1.8$\times 10^{18}$ and 2.7$\times 10^{18}$ g/s; $P_{wind}$ = 5.6$\times 10^{39}$ and 7.9$\times 10^{39}$ erg/s for $\chi$- and $\gamma$-class, respectively.
For high $kT_e$, $\dot{M}_{wind}$ = 8.2$\times 10^{17}$ and  3.0$\times 10^{18}$ g/s; $P_{wind}$ = 1.1 $\times 10^{40}$ and 4.3 $\times 10^{40}$ erg/s for $\chi$- and $\gamma$-class, respectively. Here we have considered the launching radius $R$ = 25$R_g$, where $R_g$ is the gravitational radius and mass of BH of GRS 1915+105 \cite[][]{Reid-etal2014} is $\sim$12M$_\odot$.

The total luminosity  for $\chi$- and $\gamma$-class is  6.5$\times$10$^{38}$, 1.7$\times$10$^{39}$ $erg/s$,  respectively (\citealp{Zdziarski-etal2001}), however the intensity associated to the HEP-tail in $\chi$-class is almost 35 times smaller than the total luminosity while almost 2 order smaller in case of $\gamma$-class.   
Hence, the kinetic power of wind $P_{wind}$ is greater than the luminosity of HEP-tail. As the $P_{wind}$ is calculated for $R=25R_g$ or inner-region of the disk, the HEP-tail can be generated at the inner disk region. 
In addition, the HEP-tail flux of $\gamma$-class is almost 5 times fainter than $\chi$-class (see Figure \ref{result-compare1}). And we find that the radio flux of  $\gamma$-class is almost 3 times fainter than $\chi$-class for this observation (see Figure 1 (1997 panel) and Figure 2 (1999 panel) of \citealp{Rushton-etal2010}, or see Table 1 of \citealp{Zdziarski-etal2001}; see also \cite[][]{Muno-etal2001, Fender-etal1999}).
With this similarity %(which is also true in general)
of HEP-tail and radio flux  variation between these two classes, we stress that the HEP-tail flux would be generated by
the same population of electrons that generate radio emission. In other words the location of emission region of HEP-tail and radio is same on the disk.
However, for concreteness of this result we have to consider a velocity distribution for bulk motion along with an appropriate magnetic configuration, which we will do in the future work. 
%%% $\gamma$class (1-40keV) lum = 70 unit; (40-240keV) lum = 0.6unit
%%% $\chi$class (1-80kV) lum= 40unit; (80-280) lum = 1 unit

\section{Summary}
As it was noticed earlier that a spherically diverging outflow geometry is not
a favourable one to generate the high energy power-law tail by bulk
Comptonization process \cite[e.g.][]{Laurent-Titarchuk2007}, we investigate it with  different outflow geometries mainly
a collimated  (of angle $\theta_b$) and conical outflow (of opening angle $\theta_b$) from the accretion disk.
The motivation for considering these outflow geometries is the similarity between the variation of HEP-tail flux in different spectral states and the variation of radio flux,
%due to a fact that the variation of HEP-tail flux in different spectral states is similar to the radio flux variation,
e.g., both  the radio and HEP-tail fluxes decrease when source moves from SPL to HS state,
and the radio emission generates by synchrotron emitting outflowing relativistic electron.
In XRBs, the HEP-tail is identified in high energy bin ($ E > 50$ keV) with photon index $\Gamma > 2$ and energy cut-off $E_c >200$ keV.
We have simulated the emergent BMC spectrum by a Monte Carlo scheme, considering,
the seed photon source is on the equatorial plane of the disk and inside
the outflow region.  
For conical outflow, we have considered two different outflow directions for the electrons,
in case I it is in any one of the direction along the surface
of the cone and in case II it is in any one of
direction within the conic region. In appendix head-on, we have obtained an approximate expression for $\Delta E$ (the average energy exchange per scattering) for  both collimated and  conical outflow geometries for a first and second scattering separately  in which the second scattering expression exhibits a general scenario. We have found that in second scattering the $\Delta E$ is positive for both geometries, but its magnitude is almost zero for collimated flow, while for conical flow its magnitude increases with increasing $\theta_b$.

For a low medium temperature, the HEP-tail with observed range of $\Gamma$ and $E_c$ can not be generated in collimated outflow for any $\theta_b$, but it can form in conical outflow when 
$\theta_b$ is greater than $\sim$ 30 degrees. 
At high medium temperature ($kT_e >30$ keV), the HEP-tail even generates in TC dominated spectrum, and it also finds in BMC spectrum either with collimated outflow or conical outflow for $\theta_b < 30$ degrees but this BMC spectrum is softer than the TC dominated one. 
Particularly, in conical outflow, 
in general, the emergent spectrum of case I is harder than the case II, the $\Gamma$ decreases with increasing $\theta_b$.
The BMC spectrum degenerates over parameters $kT_e$, $u_b$, $\tau$ (see equation (\ref{delta_E_bmc})) and also $\theta_b$ (for conical outflow).
For a fixed $\Gamma \sim$ 2.5, we have explored the range of degenerate parameters and summarised it in Table \ref{obs-comparison}. 

%For high $kT_e$ ($>$ 25 keV), the power-law tail with observed $\Gamma$ and $E_c$ can be generated by thermal or bulk Comptonization process. 
%In other words, for conical outflow of $\theta_b$ $<$ 30 degree and for collimated outflow, the high-energies power-law tail would be also generated, when it will found in thermal dominated case.

In Figure \ref{result-compare1} we have qualitatively described the observed HEP-tail during $\chi$- (LH) and $\gamma$-class (HS) of GRS 1915+105 for two values of $kT_e$ = 100 and 3 keV (see \citealp{Zdziarski-etal2001} for the parameters value for corresponding observation ID).   
We have found that for both values of $kT_e$ the HEP-tail of either $\chi$- or $\gamma$-class is well describe by BMC spectrum with conical outflow, and estimated $u_b$ is relativistic. In addition, the requirement that the kinetic power of wind should be greater than the total luminosity of HEP-tail is also achieved at launching radius $\sim 25R_g$. Hence the HEP-tail can be generated at inner region of the disk. Combining  this result and observation features (meant, the HEP-tail flux changes in a similar way as the radio flux from one spectral state to another) we stress that the HEP-tail and radio flux can be generated at the same region with same relativistic electron distribution.
Moreover, for few sources the polarization properties of HEP-tail has been measured (\citealp[e.g.,][]{Rodriguez-etal2015}, \citealp[for crab pulsar e.g.,][]{Vadawale-etal2017}). Therefore, the present degeneracy in BMC spectrum may be lifted out by estimating its polarization properties, we leave such study for %the polarization properties
 future work.

\section*{Acknowledgement}
%\begin{acknowledgements}
%Author is thankful to University Grants Commission, New Delhi India for providing Dr. D.S. Kothari Post-Doctoral Fellowship (201718-PH/17-18/0013).
Author thanks the anonymous referee for their comments and suggestion that have improved the presentation of the paper.
Author is supported by University Grants Commission, New Delhi India through  Dr. D.S. Kothari Post-Doctoral Fellowship (201718-PH/17-18/0013).
Author acknowledges partial financial support from Indian Space Research Organisation
(ISRO) with research Grant No.
ISTC/PPH/BMP/0362. He wishes to thank Ranjeev Misra for valuable comments on
this project and Banibrata Mukhopadhyay for their valuable
suggestions and comments over the manuscript.
%\end{acknowledgements}

%%
%v1 4 figures, 1 table; v2 5 figures, 1 table, observation comparison added and discussed, references removed/ added    
%%

\appendix
%%\begin{appendix}
\section{Monte Carlo Method}\label{MC-method}

%\bibitem[\protect\citeauthoryear{{Vadawale} et~al.}{2017}]{Vadawale-etal2017}
%Vadawale S.V., et al., 2017, \mn@doi[\natas]{10.1038/s41550-01polz_compt_pa-back20jun.tex7-0293-z}, \href
%{https://www.nature.com/articles/s41550-017-0293-z}{1-6}

%\label{lastpage}

As discussed, we considered a torus shaped  Comptonizing medium  (see Figure \ref{geo-outflow}), in which  $\tau$ is defined in the vertical direction. The electron density $n_e$ inside the medium is $n_e = \frac{\tau}{L\sigma_T}$, here $\sigma_T$ is the Thomson cross section. 
In BMC process the electrons possess a bulk velocity in addition to the thermal velocity. That is, electron's velocity has two parts, one is along the bulk direction and another one is random oriented thermal velocity, note: in special case if the thermal velocity always lies on perpendicular plane to the bulk direction then the electron has two velocity components.
In moving media the mean free path of the photons $\lambda$  exceeds that  in static medium, and it is taken into  account  \cite[e.g.,][]{Sazonov-Sunyaev2000}.
We compute the BMC spectrum using the MC method.
The algorithm for MC code is similar to that of \citet{Kumar-Misra2016} for a
thermal Comptonization and we have extended it for the BMC process by including  
 the bulk  (outflow)  motion as discussed by \citet{Laurent-Titarchuk1999, Niedzwiecki-Zdziarski2006} by neglecting general relativistic effects.
Generally, in MC calculation, a photon is tracked till it leaves the medium after either single or multiple or without scattering. This process is repeated for a large number of photons to obtain the statistically  significant results, like an emergent spectrum.  The important steps involved in  MC calculation are described as follows.

\begin{itemize}
\item In first step, we obtain the incident photon's energy $E = h\nu$ from a black body distribution at temperature $T_b$, and the electron's thermal velocity from the velocity distribution of temperature $T_e$. We fix the corona boundary of width $W$ and height $L$ at radius $r$ as shown in Figure \ref{geo-outflow}. We consider an isotropic distribution for the photons and electron's thermal velocity direction inside the corona.
  We determine the outflow direction of electron as describe in Figure \ref{geo-outflow}, and take electron's bulk speed $u_b$.
The $\lambda$ for a  photon of energy $E$ is computed for a given $n_e$ (or $\tau$), $T_e$, and $u_b$. We compute the above quantities in the global coordinate fixed at centre of torus (or at the compact object).

\item Next, we obtain the collision free path of the photon $l_f$ inside the corona using an exponential pdf (probability distribution function), $\exp\left(\frac{-l_f}{\lambda}\right)$, and find the scattering point P position in the global coordinate. Further we determine the condition for occurrence of scattering at P. If the point P is inside the corona then the scattering will happen otherwise the photon will escape the corona without scattering.
 
\item In second step we transform the quantities from global coordinate to the comoving frame of electron's bulk velocity. By knowing the angle between photon's direction and electron's bulk velocity, we determine the Doppler boosted frequency and aberration for photon in the comoving frame.

\item In third step to describe the Compton scattering, we transform the quantities from comoving frame to the electron rest frame (as in comoving  frame, the electron has only thermal velocity). 
  By knowing the angle between photon's direction and electron's thermal velocity  in comoving frame, we determine the Doppler boosted frequency and aberration for photon in the rest frame.

\item Using Klein-Nishina cross section \\$\Big(= \frac{1}{4}r_o^2\left(\frac{k'}{k}\right)^2  \left[\frac{k}{k'}+\frac{k'}{k}-\sin^2\theta\right]$, here $k = \frac{h\nu}{c}$ is the incident photon momentum, $k' = \frac{h\nu'}{c}$ is the scattered photon momentum, $\nu$ and $\nu'$ are incident and scattered photon frequency, respectively, $r_o$ is the classical radius of the electron, $\theta$ is the scattering angle\big), we extract the scattering angle. Using Compton scattering frequency formula $\Big(\frac{\nu'}{\nu} = \frac{1}{1+\frac{h\nu}{m_e c^2}(1-\cos\theta)}\Big)$ in the electron rest frame we compute the scattered photon frequency. 

\item Next we find the scattered photon's direction in electron rest frame. We transform back the quantities (frequency of scattered photon and its aberration)  from electron rest frame to the comoving frame. We find the scattered photon's direction in comoving frame. Finally we compute the scattered photon's frequency and its direction in the global coordinate by transforming back the quantities from comoving frame to the global coordinate (or lab frame).

\item Next we compute the collision free distance $l_f$ for the scattered photon (of energy $E' = h\nu')$ and find the next scattering point $P'$ position. If the $P'$ lies inside the corona then next scattering will occur otherwise photon will escape the medium.

\item For next scattering we proceed the calculations with treating $k'$ of previous scattering as an incident photon, and follow the same steps %(described above)
  till the scattered photon escapes the medium.

\end{itemize}

\subsection{MC results verification}
\begin{figure}
\centering%$

\includegraphics[width=0.5\textwidth]{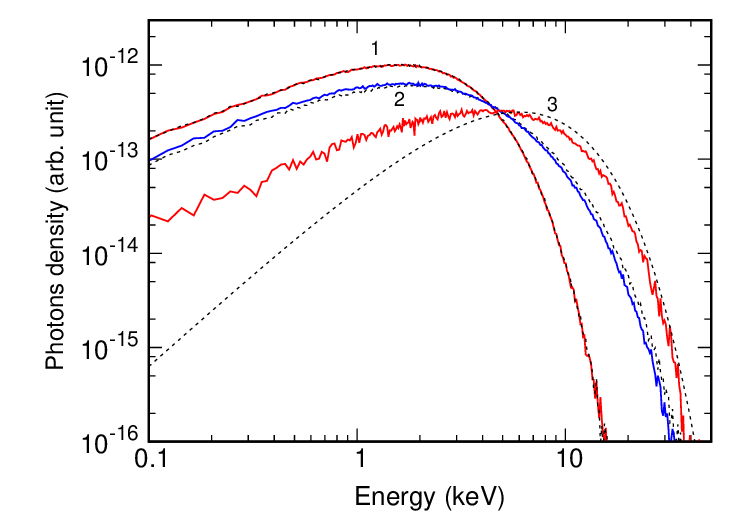} 

\caption{Comparison of the BMC and TC spectra for  Monte Carlo results verification. %Spectra comparing Monte Carlo results. 
The solid  curves are for bulk Comptonization ($kT_e$ = 2 keV, $u_b$=0.0766c) 
and dashed  curves are for thermal Comptonization ($kT_e$ = 3 keV, $u_b$=0.0).
The curves 1, 2, and 3 are
for single scattering, multiple scattering ($\langle N_{sc} \rangle$ $\sim$
46), and Wien peak ($\langle N_{sc} \rangle$ $>$ 500), respectively.  Here, the dashed curve 3 is $ \propto E^2  \exp \big( \frac{-E}{kT_e} \big)$, which also determines, in general, the exponential high energy cut-off ($E_c$) for photons density in  non-relativistic TC process.
} 
\label{chk-delE}
\end{figure}

As discussed in the method section \S\ref{sec:cal-res} we check here the MC code by comparing
the simulated bulk Comptonized spectrum to thermal
Comptonized one for those parameters sets which have same
$\Delta E$, for example, the sets ($kT_{e}$ = 3 keV,  $u_b$ = 0) for TC and ($kT_{e}$ = 2 keV,  $u_b$ = 0.0766c) for BMC; here we neglect the term $\frac{4u_b}{c\tau}$ in calculation of $\Delta E$ (i.e., we consider only the random Compton scattering, see equation (\ref{delta_E_bmc})). 
The emergent spectra for both sets are identical for either single, or
multiple scattering or Wien peak spectrum (i.e. for large
$\langle N_{sc} \rangle$ $\sim$ 500) which are shown by the curves 1, 2 and 3 in Figure \ref{chk-delE}, respectively.

%\section{Approximate first order (in $u_b$) term of $\Delta E$}\label{head-on}
\begin{small}\section{Approximate $\Delta E$ for collimated $\&$ conical outflow}\label{head-on}\end{small}
%{Interpretation of first order in $u_b$ term of $\Delta E$}\label{head-on}

In this Appendix we obtain the approximate expression for $\Delta E$ by using the Compton scattering frequency formula for  both outflow geometries, collimated and conical. For this, %to avoid the complexity
  we  compute the $\Delta E$ for every  scattering, for simplicity, up to  the second scattering. %, e.g., for the second scattering we consider the first scattered photon as an incident photon.
  In first scattering the directions of incident photon and electron bulk velocity are fixed, %(i.e., $\alpha_i  = \theta_b$ for both outflow geometries),
  so the $\Delta E$ of first scattering will contain dominantly only the first order in $u_b$ term.  In second scattering case the directions of incident photon (or first scattered photon) can be any, so for the considered bulk velocity geometry, it also demonstrates  a general situation. %, and one does not need to compute $\Delta E$ for further scattering.
  As the scattering planes are isotropically distributed, we formulate $\Delta E$ for a fixed scattering plane. % this results will valid in general.  
 In addition, we perform the calculation in Thomson regime \Big($\frac{h\nu}{\gamma m_e c^2} \ll 1$\big), where %and in this regime
  the Klein-Nishina differential cross section is  $\frac{d\sigma}{d\Omega} = \frac{r_o^2}{4}\left(1+\cos^2\theta\right)$. %, where $r_o$ is the classical radius of the electron.   

In lab frame the Compton scattering frequency formula  for the incident photon of energy E ($=h\nu$) and the incident electron of velocity  $u_b$ is written as
  \[ \frac{\nu'}{\nu} = \frac{1 - \frac{u_b}{c}\cos \alpha}{1 - \frac{u_b}{c}\cos \alpha' + \frac{h\nu}{\gamma m_e c^2}(1-\cos \theta)} \]
  where  $\alpha$ is the angle between incident photon and incident electron,  $\alpha'$ is the angle between scattered photon and incident electron, and $\gamma$ = $\left(\sqrt{1-\frac{u_b^2}{c^2}}\right)^{-1}$. In general, for $i^{th}$ scattering the $\Delta E_i$ for a low energy photon in Thomson  regime is
    %In addition, for the relativistic electron the Klein-Nishina cross section predicts that the scattered photon will lie along the incident electron velocity direction, i.e., $ \alpha' = 0$.}
    \begin{equation}\label{eq:appen-de}  \frac{ \Delta E_i}{E_i} =  \frac{ \nu'_i- \nu_i}{ \nu_i} =  \frac{\frac{u_b}{c}(\cos \alpha'_i-\cos\alpha_i)}{1-\frac{u_b}{c}\cos \alpha'_i}  \end{equation}
    here, the variable with  subscript $i$ is related to the i$^{th}$ scattering. For example, a) for the head-on scattering the $ \alpha_1$ is  $\pi$ and $\frac{ \Delta E_1}{E_1}$ is $=\frac{\frac{u_b}{c}(1+\cos \alpha'_1)}{1-\frac{u_b}{c}\cos \alpha'_1} (> 0)$; b) for the rear-end scattering $\frac{ \Delta E_1}{E_1}$ = $\dfrac{\frac{u_b}{c}(\cos \alpha'_1-1)}{1-\frac{u_b}{c}\cos \alpha'_1}  (< 0)$, here $ \alpha_1 =  0$. Hence in the Comptonizing medium, where the photon suffers mostly head-on scattering its energy after scattering will increase, while the opposite is true for the rear-end scattering. Therefore, the first order term in the electron bulk velocity of $ \Delta E $ (in equation  \ref{delta_E_bmc}) is contributed due to the mainly head-on scattering. 

For a given scattering plane, $\alpha'_i = \alpha_i \pm \theta_i $,  so for a given $\alpha_i$ the $\alpha'_i$ is function of only the $\theta_i$. The averaged $\cos\alpha'_i$ over $\theta_i$ can be computed as $\langle\cos\alpha'_i\rangle$ = $\frac{\int_0^\pi (1 + \cos^2\theta_i)\cos\alpha'_i d\theta_i}{\int_0^\pi (1 + \cos^2\theta_i) d\theta_i}$ (note: in plane the differential solid angle is $d\Omega = d\theta$). %Then we %qualitatively
One can estimate  the approximate averaged  $\Delta E_i$ (or $\langle \Delta E_i\rangle$) for a given $\alpha_i$ using the  equation (\ref{eq:appen-de}) %by  substituting the value of $\langle\cos\alpha'_i\rangle$ in equation \ref{eq:appen-de}.
as  $\frac{\langle \Delta E_i\rangle}{E_i}$ =  $\frac{u_b}{c}\frac{\langle\cos \alpha'_i \rangle-\cos\alpha_i}{1-\frac{u_b}{c}\langle \cos \alpha'_i\rangle}$. Here, for $i \geq 2$ the $\cos \alpha_i$ would be replaced by the averaged $\cos \alpha_i$ over the $\theta_{i-1}$ ($\langle \cos \alpha_i \rangle$), as the incident photon of  $i^{th}$ scattering is also a scattered photon  of $(i-1)^{th}$ scattering.

{\tt \textbf{First scattering:}} Since, in first scattering $\alpha_1 =  \theta_b$ for both collimated and conical flow, so their single scattered Comptonized spectrum will be same.
%  As the incident photon direction is vertically outward, so for both outflow geometries (collimated or conical flow),  for first scattering the angle $\alpha_1$ is $\theta_b$ and correspondingly  the first scattered Comptonized spectrum is identical. 
  For a given scattering plane %the angle $\alpha'_1$ be $\alpha_1 \pm \theta_1$,
%and corresponding $\frac{\Delta E}{E_1}$ are $\left. \frac{ \Delta E}{E_1}\right|_{\alpha'_1 = \theta_b \pm \theta_1}$.
%  we refer the term $\frac{\Delta E}{E_1}$ to $\left. \frac{ \Delta E}{E_1}\right|_a$ for $\alpha'_1 = \theta_b - \theta_1$ and to $\left. \frac{ \Delta E}{E_1}\right|_b$ for $\alpha'_1 = \theta_b + \theta_1$ and
%Their resultant is
%{\small $\frac{\Delta E_1}{E_1} = \frac{1}{2}\left(\left. \frac{ \Delta E_1}{E_1}\right|_{\alpha'_1 = \theta_b + \theta_1} + \left. \frac{ \Delta E_1}{E_1}\right|_{\alpha'_1 = \theta_b - \theta_1} \right)$}.  
%
% 
 % For first scattering, the  averaged $\left. \frac{ \Delta E_1}{E_1}\right|_a$ and $ \left. \frac{ \Delta E_1}{E_1}\right|_b$ are
  %\begin{small} \[\left. \frac{\langle \Delta E_1\rangle}{E_1}\right|_a = \frac{(\frac{16}{9\pi}\sin\theta_b - \cos\theta_b)\frac{u_b}{c}}{1-\frac{u_b}{c}\frac{16}{9\pi}\sin\theta_b} \ \ \left. \frac{\langle \Delta E_1\rangle}{E_1}\right|_b =\frac{(-\frac{16}{9\pi}\sin\theta_b - \cos\theta_b)\frac{u_b}{c}}{1+\frac{u_b}{c}\frac{16}{9\pi}\sin\theta_b}\]\end{small}
  the  averaged %\begin{small}
  $\left. \frac{ \Delta E_1}{E_1}\right.$ at a given $\alpha_1$  %|_{\alpha'_1 = \theta_b \pm \theta_1}$
  are  $\left. \frac{\langle \Delta E_1\rangle}{E_1}\right|_{\alpha'_1 = \theta_b \pm \theta_1} = \frac{u_b}{c}\frac{\pm\frac{16}{9\pi}\sin\theta_b - \cos\theta_b}{1\mp\frac{u_b}{c}\frac{16}{9\pi}\sin\theta_b}$, %\end{small}  
here we use the relations $\frac{\int_0^\pi (1+\cos^2\theta_1)\sin\theta_1 d\theta_1}{\int_0^\pi (1 + \cos^2\theta_1)d\theta_1}$ = $\frac{16}{9\pi}$, $\frac{\int_0^\pi (1+\cos^2\theta_1)\cos\theta_1 d\theta_1}{\int_0^\pi (1 + \cos^2\theta_1)d\theta_1}$ = 0 .
Finally, by averaging them the resultant  is
\[\begin{small} \frac{\langle \Delta E_1 \rangle}{E_1} %=\frac{1}{2}\left(\left. \frac{ \Delta E_1}{E_1}\right|_{\alpha'_1 = \theta_b + \theta_1} + \left. \frac{ \Delta E_1}{E_1}\right|_{\alpha'_1 = \theta_b - \theta_1} \right)
  = \frac{\frac{u_b}{c} (-\cos\theta_b +\left(\frac{16}{9\pi}\right)^2\frac{u_b}{c} \sin^2\theta_b)}{1-\left(\frac{u_b}{c}\frac{16}{9\pi}\sin\theta_b\right)^2} \end{small}\]
Here, we note that $\langle\Delta E_1\rangle > 0$ for $\theta_b > 80$ degrees, however from the MC calculation we find a similar trend but for $\theta_b > 50$ degrees. In addition, the $\frac{\langle \Delta E_1 \rangle}{E_1}$ is $\frac{u_b}{c}$ and $\frac{-u_b}{c}$ for head-on and rear-end scattering, respectively.

{\tt \textbf{Second scattering:}}  %The first scattered photon (of frequency $\nu'_1$ or
  As mentioned earlier, the  incident photon for second scattering %of frequency $\nu_2$, $\nu'_1$ = $\nu_2$)
  can have  any $\theta$-angle in global coordinate (or $\theta_1$), %as incident photon for first scattering is vertically upward, i.e., along the $z$-axis)), 
  so  it will see different initial collision environments in  both outflow geometries. For example, if the $\theta$-angle of $\nu_2$ photon  is $\theta_b$ then in collimated flow the $\alpha_2$ is either 0 or 2$\theta_b$ while in conical flow it has range of values, $\alpha_2 \equiv$ [0,2$\theta_b$]. In general, for a given $\theta_1$ the $\alpha_2$ is $\theta_b \pm \theta_1$ and $\equiv$[$\theta_b-\theta_1,\theta_b+\theta_1$] for collimated and  conical flow, respectively.

In collimated flow, for a given scattering plane there are four different possibilities of $\alpha'_2$ for a given $\alpha_2$, these $(\alpha_2; \alpha'_2)$ are ($\theta_b-\theta_1; \theta_b-\theta_1\pm\theta_2$), ($\theta_b+\theta_1; \theta_b+\theta_1\pm\theta_2$). and corresponding
  $\left.\frac{\langle\Delta E_2\rangle}{E_2}\right|_{\alpha'_2 =\alpha_2 \pm \theta_2}^{\alpha_2 =\theta_b - \theta_1}$ = $\frac{u_b}{c} \frac{\pm\left(\frac{16}{9\pi}\right)^2\cos\theta_b-\left(\frac{16}{9\pi}\right)\sin\theta_b}{1 \mp \frac{u_b}{c}\left(\frac{16}{9\pi}\right)^2\cos\theta_b}$, and
  %$\left.\frac{\langle\Delta E_2\rangle}{E_2}\right|_{\alpha'_2 =\alpha_2 + \theta_2}^{\alpha_2 =\theta_b - \theta_1}$ = $\frac{\left(\frac{16}{9\pi}\right)^2\cos\theta_b-\left(\frac{16}{9\pi}\right)\sin\theta_b}{1-\frac{u_b}{c}\left(\frac{16}{9\pi}\right)^2\cos\theta_b}$,
  %$\left.\frac{\langle\Delta E_2\rangle}{E_2}\right|_{\alpha'_2 =\alpha_2 - \theta_2}^{\alpha_2 =\theta_b - \theta_1}$ = $\frac{-\left(\frac{16}{9\pi}\right)^2\cos\theta_b-\left(\frac{16}{9\pi}\right)\sin\theta_b}{1+\frac{u_b}{c}\left(\frac{16}{9\pi}\right)^2\cos\theta_b}$,
 $\left.\frac{\langle\Delta E_2\rangle}{E_2}\right|_{\alpha'_2 =\alpha_2 \pm \theta_2}^{\alpha_2 =\theta_b + \theta_1}$ = $\frac{u_b}{c} \frac{\mp\left(\frac{16}{9\pi}\right)^2\cos\theta_b+\left(\frac{16}{9\pi}\right)\sin\theta_b}{1 \pm \frac{u_b}{c}\left(\frac{16}{9\pi}\right)^2\cos\theta_b}$. 
%  $\left.\frac{\langle\Delta E_2\rangle}{E_2}\right|_{\alpha'_2 =\alpha_2 + \theta_2}^{\alpha_2 =\theta_b + \theta_1}$ = $\frac{-\left(\frac{16}{9\pi}\right)^2\cos\theta_b+\left(\frac{16}{9\pi}\right)\sin\theta_b}{1+\frac{u_b}{c}\left(\frac{16}{9\pi}\right)^2\cos\theta_b}$
%  and $\left.\frac{\langle\Delta E_2\rangle}{E_2}\right|_{\alpha'_2 =\alpha_2 - \theta_2}^{\alpha_2 =\theta_b + \theta_1}$ = $\frac{\left(\frac{16}{9\pi}\right)^2\cos\theta_b+\left(\frac{16}{9\pi}\right)\sin\theta_b}{1-\frac{u_b}{c}\left(\frac{16}{9\pi}\right)^2\cos\theta_b}$ 
The resultant $\frac{\langle\Delta E_2\rangle}{E_2}$ is the averaged of these four quantities. We note that the resultant is almost zero and positive, and we obtain the same results by employing the MC method.   %%%%%results show same.

In conical flow, as mentioned earlier,  there is a range of $\alpha_2$ for a given  $\theta_1$, specifically the $\phi$-plane of electron bulk velocity makes angle from 0 to $\pi$ with $\phi$-plane of $\nu'_1$ photon. To compute the $\Delta E$ we consider a $u_b$ whose $\phi$-plane makes angle $x$ with $\phi$-plane of $\nu'_1$ photon, so   $\cos\alpha_2$ = $\cos(\theta_b \pm \theta_1)\cos\theta_b +\sin((\theta_b \pm \theta_1)\sin\theta_b\cos x$ for $\alpha'_1$ = $\theta_b \pm \theta_1$. As $\alpha'_2$ = $\alpha_2 \pm \theta_2$, the different possible averaged values of $\Delta E$ are
$\left.\frac{\langle\Delta E_2\rangle}{E_2}\right|_{\alpha'_2 =\alpha_2 \pm \theta_2}^{\alpha_2 =\theta_b - \theta_1}$ =$\frac{u_b}{c} \frac{\mp P - \frac{8}{9\pi}\sin(2\theta_b)(1-\cos x)}{1\pm  \frac{u_b}{c} P}$, and
$\left.\frac{\langle\Delta E_2\rangle}{E_2}\right|_{\alpha'_2 =\alpha_2 \pm \theta_2}^{\alpha_2 =\theta_b + \theta_1}$ =$\frac{u_b}{c} \frac{\mp Q - \frac{8}{9\pi}\sin(2\theta_b)(-1+\cos x)}{1\pm  \frac{u_b}{c} Q}$, where

\noindent  {\small $P$ =} {\small  $\frac{16}{9\pi}\left(1-\frac{y}{4}\left( (1 + \frac{\cos(2\theta_b)}{6}) \cos^2\theta_b \right.\right.$  $\left.\left.+ (1 - \frac{\cos(2\theta_b)}{6}) \sin^2\theta_b\cos^2x \right.\right.$  $\left.\left.\pm \frac{1}{6}\sin^2(2\theta_b)\cos x\right)\right)$ and}

\noindent  {\small  $Q$ = $\frac{16}{9\pi}\left(1-\frac{y}{4}\left( (1 + \frac{\cos(2\theta_b)}{6}) \cos^2\theta_b \right.\right.$  $\left.\left.+ (1 - \frac{\cos(2\theta_b)}{6}) \sin^2\theta_b\cos^2x \right.\right.$  $\left.\left. + \frac{1}{6}\sin^2(2\theta_b)\cos x\right)\right)$.} 
  The positive and negative  sign of forth right hand side of P are for $\theta_b < \theta_1$ and $\theta_b > \theta_1$, respectively. Here, for simplicity we consider $\sin\alpha_2 \approx (1 - \frac{y}{2}\cos^2\alpha_2)$, for $y$ = 1 it is a first two terms of binomial series of $\sin\alpha_2$, however in present context (or in crude approximation) the  $y \sim$2 well describes the $\Delta E$  in the [0,$\pi$] range, even in the range of [0,2$\pi$] as $\langle  \cos\alpha'_2\rangle$ = $\mp\frac{16}{9\pi}\langle\sin\alpha_2\rangle$. The resultant $\frac{\langle\Delta E_2\rangle}{E_2}$ is the averaged of these four quantities. We find that the resultant is positive for any $\theta_b$ and its magnitude increases with increasing $\theta_b$, and qualitatively, this trend does not depend on the $y$.
The  MC calculation is consistent with this result.
  In summary, for second scattering (which reveals also a general situation), the  resultant $\frac{\langle\Delta E_2\rangle}{E_2}$ is positive for both geometries, but its magnitude is almost zero for collimated flow, while for conical flow its magnitude increases with increasing $\theta_b$. %$\approx$ 0 in collimated flow  and in conical flow  $\frac{\langle\Delta E_2\rangle}{E_2}$ $>$ 0
  %
  %two extreme values of $\alpha_2$, $\theta_b -\theta_1 + p$ and $\theta_b +\theta_1 - p$, here $p$ is the increment in $\alpha_2$  from the $\phi$-plane of first scattered plane, and it ranges from 0 to $\cos^{-1}[(\cos(\theta_b -\theta_1)\cos(\theta_b)]$, in which the . 

%similar to the first scattering case, to compute the $\langle\Delta E_2\rangle$ we also n the averaged $\cos\alpha_2$ 

%\mg{\tt {Next/ second scattering in conical flow:}
%}
 %%  \end{appendix}
 \def\aap{A\&A}%
\def\aapr{A\&A~Rev.}%
\def\aaps{A\&AS}%
\def\aj{AJ}%
\def\actaa{Acta Astron.}%
\def\araa{ARA\&A}%
\def\apj{ApJ}%
\def\apjl{ApJ}%
\def\apjs{ApJS}%
\def\apspr{Astrophys.~Space~Phys.~Res.}%
\def\ao{Appl.~Opt.}%
\def\aplett{Astrophys.~Lett.}%
\def\apss{Ap\&SS}%
\def\azh{AZh}%
\def\bain{Bull.~Astron.~Inst.~Netherlands}%
\def\baas{BAAS}%
\def\bac{Bull. astr. Inst. Czechosl.}%
\def\caa{Chinese Astron. Astrophys.}%
\def\cjaa{Chinese J. Astron. Astrophys.}%
\def\fcp{Fund.~Cosmic~Phys.}%
\def\gafd{Geophys.\ Astrophys.\ Fluid Dyn.}
\def\gca{Geochim.~Cosmochim.~Acta}%
\def\grl{Geophys.~Res.~Lett.}%
\def\iaucirc{IAU~Circ.}%
\def\icarus{Icarus}%
\def\jcap{J. Cosmology Astropart. Phys.}%
\def\jcp{J.~Chem.~Phys.}%
\def\jfm{JFM}
\def\jgr{J.~Geophys.~Res.}%
\def\jqsrt{J.~Quant.~Spec.~Radiat.~Transf.}%
\def\jrasc{JRASC}%
\def\mnras{MNRAS}%
\def\memras{MmRAS}%
\def\memsai{Mem.~Soc.~Astron.~Italiana}%
\def\na{New A}%
\def\nar{New A Rev.}%
\def\nat{Nature}%
\def\natas{Nature Astronomy}%
\def\nphysa{Nucl.~Phys.~A}%
\def\pasa{PASA}%
\def\pasj{PASJ}%
\def\pasp{PASP}%
\def\physrep{Phys.~Rep.}%
\def\physscr{Phys.~Scr}%
\def\planss{Planet.~Space~Sci.}%
\def\pra{Phys.~Rev.~A}%
\def\prb{Phys.~Rev.~B}%
\def\prc{Phys.~Rev.~C}%
\def\prd{Phys.~Rev.~D}%
\def\pre{Phys.~Rev.~E}%
\def\prl{Phys.~Rev.~Lett.}%
\def\procspie{Proc.~SPIE}%
\def\qjras{QJRAS}%
\def\rmxaa{Rev. Mexicana Astron. Astrofis.}%
\def\sgg{Stud.\ Geoph.\ et\ Geod.}
\def\skytel{S\&T}%
\def\solphys{Sol.~Phys.}%
\def\sovast{Soviet~Ast.}%
\def\ssr{Space~Sci.~Rev.}%
\def\zap{ZAp}%
\def\memsai{Memorie della Societa Astronomica Italiana}

%%%%%%%%%%%%%%%%%%%%%%%%%%%%%%%%%%%%%%%%%%%%%%%%%%
\section*{Data Availability}
In the Figure
 \ref{result-compare1}, the data is taken from the published work \citealp{Zdziarski-etal2001}.



\bibliographystyle{elsarticle-harv}
\bibliography{pap4}

\end{document}